# Adaptive Partitioning and its Applicability to a Highly Scalable and Available Geo-Spatial Indexing Solution

David W. LeJeune Jr.

dlejeune@stopllc.com

**Abstract**

Satellite Tracking of People (STOP) tracks thousands of GPS-enabled devices 24 hours a day and 365 days a year. With locations captured for each device every minute, STOP servers receive tens of millions of points each day. In addition to cataloging these points in real-time, STOP must also respond to questions from customers such as, *"What devices of mine were at this location two months ago?"* They often then broaden their question to one such as, *"Which of my devices have ever been at this location?"* The processing requirements necessary to answer these questions while continuing to process inbound data in real-time is non-trivial.

To meet this demand, STOP developed Adaptive Partitioning to provide a cost-effective and highly available hardware platform for the geographical and time-spatial indexing capabilities necessary for responding to customer data requests while continuing to catalog inbound data in real-time.

## Introduction

A weakness with conventional partitioning is its static nature. Decisions made at the outset of a project often need to be reconsidered as it matures. Moreover, while selecting an inefficient partitioning approach can dramatically impact performance, finding a "correct" way to partition data is not always apparent given that complex query and load requirements often conflict with one another [RZL02]. When projects do require a new partitioning scheme, updating midstream can be expensive. Re-partitioning a highly transactional and large dataset requires planning, additional processing and can result in downtime of the affected data structures.

The even distribution of data across a partition does not necessarily constitute an effective partitioning strategy. One must also consider the elimination of processing hot spots [SCSVR 08]. The reduction of hot spots by transaction distribution is an effective way to mitigate their impact [NDV 03]. Most relational database vendors today offer parallel options that can address this distribution of processing by *scaling-out* horizontally in a multi-node shared-nothing architecture [PRMSDPS 09]. While it has been argued that shared-nothing architectures are superior for scaling horizontally [S86], some of the more established RDBMS vendors, including Oracle and Microsoft (with Exadata and Madison/SQL Server products, respectively), have just begun offering shared-nothing parallel database solutions [PRMSDPS 09].

Others argue that parallel databases do not always provide the best solution for accommodating data sets with particularly large sizes and transaction throughput. Those in the NoSQL movement believe there is a simpler, more efficient and cost-effective approach that does not rely on single "monolithic" parallel database architecture; they instead support an open source, distributed, non-relational approach [F 10]. Traditional relational database capabilities such as feature-rich SQL or transaction consistency are relaxed, or in some cases abandoned, in favor of independent locally sufficient data stores that are distributed (in many cases geographically), across low-end commodity servers.

NoSQL proponents also believe the rate of data production is already outpacing Moore's law [HAMS 08]. This, coupled with an increasing cost benefit for low-end "commodity" servers, will drive the size of distributed computing into the 1,000 and 10,000 node range [AKDASR 09]. They point to implementations such as Google's BigTable [CGHWBCG 08], Facebook's Cassandra [LM 09], Yahoo's PNUTS, [CRSSBJPWY 08] and Amazon's Dynamo [DHJKLPSVV 07] that have abandoned traditional RDBMS architectures in favor of a NoSQL paradigm that provides an ability to scale horizontally in a non-heterogeneous shared-nothing architecture [AKDASR 09].

While those on both sides agree that the explosion of online transaction processing has driven a need for horizontally scalable database solutions, there is considerable debate as to which architecture is best suited to handle this growth in "big data". Some feel the NoSQL movement simply chooses to ignore the capabilities of existing modern parallel database solutions [ALTFFP 10]. NoSQL proponents contend that parallel databases are more difficult to establish and maintain [PRMSDPS 09] and are more susceptible to hardware failures [PRMSDPS 09]. With pros and cons to each approach, the selected architecture is often based on the type of problem being solved.

STOP's design for Adaptive Partitioning benefited from NoSQL concepts and implementations. When weighing a relational parallel database versus a NoSQL approach, the benefits gained from using the distributed, shared-nothing data store residing on locally sufficient, low-end commodity servers in multiple geographic locations outweighed the drawbacks associated with moving away from a more traditional relational database.

**Deciding Factors for Adaptive Partitions**

STOP conceived the idea for Adaptive Partitioning in part to address transactional latency experienced with production track point data store (at the time Oracle 10g Enterprise using the Partitioning option). This latency became more apparent when the database attempted to service more complicated data requests in parallel with cataloging inbound data in real-time. These data requests typically involved data spread across a large range in the data store that did not conform to the original partitioning scheme. Regularly measured in seconds or in some cases minutes, latency in these transactions typically resulted from the database attempting to maintain a consistent data state while determining the query's results.

The original data store used a range partition based on the month the track data was reported as well as a sub-hash partition on the providing device's unique identifier. During the first eighteen months the system's data grew exponentially. The types of data requests also became more computationally intense. For example, instead of requesting information for a single device, users instead requested points for all their devices for a given geographic location across multiple months (i.e., partitions). Hot spots in the data also developed as users requested more recent data. All of these factors forced a rethinking of the partitioning strategy that might result in less dense partitions and at the same time address hot spots in the data.

While problematic, the explosive growth of device traffic and user's appetite for data did provide observations that assisted in the design of Adaptive Partitioning. For example, system devices do not update data they have already reported, nor are their track data removed from the data store aside from periodic historical archiving. The knowledge that data would not be modified after cataloging allowed the design to use an eventually consistent data transaction model critical for the design of Adaptive Partitions.

Relational databases must support a wide range of functionality. Previous "must-have" features in traditional OLTP databases such as logging, locking (i.e., for a two-phase commit), and multi-threaded writes are no longer warranted for every application due to the overhead they incur [HAMS 08]. The highly transactional portion of the STOP application that handles track point data could execute without some of these features. For example, database logging for the purpose of data recovery was negated by keeping redundant copies of the data stored elsewhere. Because data requests could be satisfied using an eventual consistency model, there was no need for record locking and buffer management. Distributing the partition's data across multiple low-end "commodity" servers avoided the need for multiple writing threads when cataloging the data.

Although it is significant in parole and probation offender monitoring space, STOP's transaction rates pale in comparison to what is handled in other online spaces. However, it can still be a challenge to scale processing horizontally in a cost-effective manner. Further, STOP lacks the resources that a Google, Amazon, FaceBook, or Microsoft has to tackle this problem. It is the well-defined and compartmentalized nature of STOP's transaction set that led to the idea of Adaptive Partitioning. While its design is more limited in scope than solutions implemented by other larger companies, it is built on the same shared-nothing distributed processing principles allowing for horizontal growth across inexpensive hardware in multiple geographic locations.

**Adaptive Partitioning Explained**

The primary performance goal for Adaptive Partitioning was to keep cataloged data evenly distributed across tables in the partition while eliminating processing hot spots created by user data requests. Data cataloging within an Adaptive Partition is spread across multiple low-end "commodity" servers that can reside in multiple geographic locations. Tables that become "too hot" from a transaction processing perspective are split, which accommodates the higher transaction rate. User data requests are satisfied by querying the affected tables in parallel, collating the results and returning them to the end user.

Data distribution within an Adaptive Partition is accomplished by using a range-based consistent hashing approach [KLLLLP 97]. While there were initial concerns that issues such as skewing of data [DG 92] might prevent the efficient distribution of data and the associated transactions across partitions, these issues did not affect data distribution. Even distribution results as the partition adapts over time.

Each Adaptive Partition has a main index, or map, that describes how data is distributed across the partition. Inbound requests to catalog new data or retrieve existing data are resolved against the main index to determine the table(s) in the partition that should service the request. As the characteristics of the cataloged data change, the main index is updated to allow subsequent data to be more evenly distributed.

For cataloging new data, an Adaptive Partition's main index differentiates between tables that are live and are actively having data written to them versus those that were closed and are available solely for servicing queries. To ensure a consistent update of the main index across multiple servers, changes are made to take effect at a predetermined future date, thereby allowing time for the changes to be replicated to all servers that service the partition. Updates include adding references to new live tables as well as closing existing tables that are being superseded by the new tables.

An Adaptive Partition's live tables are closed as a result of two conditions. First, when the data range associated with a particular table is experiencing a higher than predetermined "optimum" transaction rate, the range is split and new tables are created to accommodate new distribution. Second, if a table has not experienced a transaction rate that exceeds the optimum but the table has been active long enough to warrant a redistribution of its load, it is closed and replaced with a new table that supports the cataloging of an identical range of data. Ideally, tables are closed when they are approaching or are at their "optimum" transaction rate.

In the second case, tables are closed to simplify the consistency model. Because the data itself is static (i.e., devices do not update data they have already reported), a closed table can exclusively service queries and thus remain in a consistent state. Transaction consistency is not required when servicing requests for data whose results span multiple tables. Where other solutions such as Yahoo's data serving platform PNUTS [CRSSBJPWY 08] expose the complexity of versioning and maintaining transaction consistency via API's when accessing their data store, Adaptive Partitions instead implement an eventually consistent model where the data that is returned satisfies the request at the time of execution against a particular table in the partition.

Two settings within an Adaptive Partition dictate how its main index will update over time. An optimum record count or threshold determines when tables have become too dense and should be split to better distribute and handle cataloging data in the future. Splitting the data range also results in a more evenly distributed load as tables that are more active can more quickly split and distribute their transactions to the tables that replace them. If tables do not exceed this maximum record threshold, a second setting indicates a maximum amount of time a table will remain live before it is closed and a new table is created to take its place.

Figure 1 illustrates a scenario of cataloging data based on the first letter of an alphanumeric text string. The main index is periodically evaluated and splits when data in tables exceeds the optimum number of records per table (i.e., the first setting within the Adaptive Partition). Two new live tables (Tables 2 and 3) are created and the prior table (Table 1) is closed. A time-spatial range (Times 0 – 1 for Table 1) is associated with closed tables that allow them to service requests for historical data.

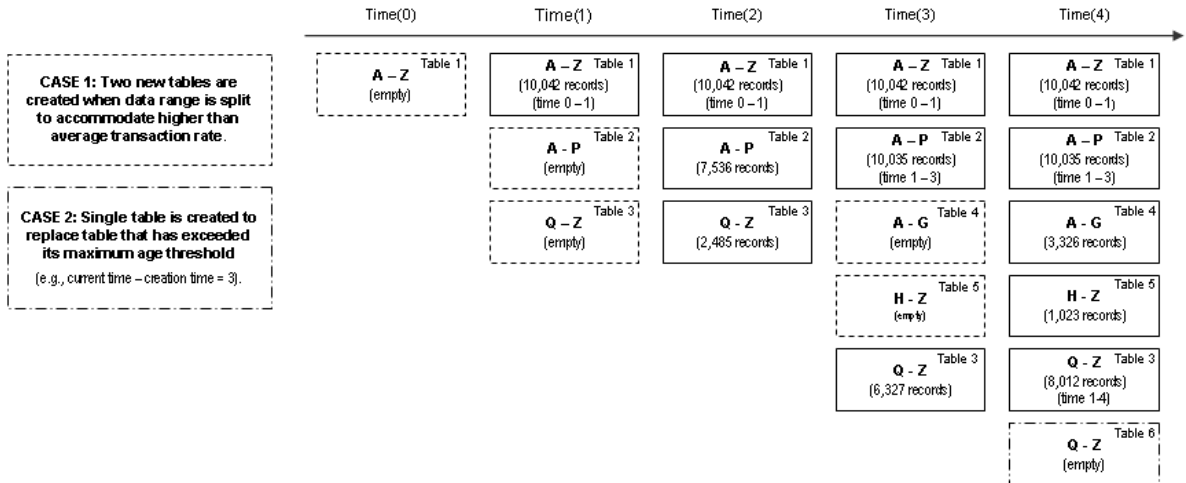

**Figure 1: Evolution of an Adaptive Partition**

The main index is also evaluated periodically for tables that have outlived their maximum life expectancy (i.e., the second setting within the Adaptive Partition). Figure 1 depicts this scenario as three units of time. When this condition is detected, the table (Table 3) is closed and assigned a time-spatial range (Times, 1-4), and a single new live table (Table 6) is created in its place.

Although it is used by Adaptive Partitioning, consistent hashing in a shared-nothing horizontally scalable architecture is not a new technique. Google's Bigtable, [CGHWBCG 08], Amazon's Dynamo, [DHJKLPSVV 07] and Yahoo's PNUTS [CRSSBJPWY 08] have all developed systems that re-distribute data once certain pre-determined thresholds are met. STOP's simplified solution of dynamically updating the partition's main index and thereby its hashing algorithm does offer some inherent advantages. For example, with their ordered-table solution PNUTS, Yahoo employs a pre-planning phase [SCSVR 08] whereby data being cataloged is staged so that an optimum bulk insertion distribution that helps eliminate hot spots can be achieved. Because Adaptive Partitions allows the redistribution of inbound data over time, pre-planning a staging of data is no longer necessary.

**Clustering Adaptive Partitions**

Clustering of an Adaptive Partition can be broken down into three functions. First, its main index is synchronized across on all servers in the cluster, thus allowing all nodes within the cluster to handle inbound requests simultaneously. Second, the cluster constantly monitors current data and transaction distribution levels and accommodates new servers coming online by including them in future updates to its main index.

Finally, multiple copies of the data are cataloged across servers and locations, allowing the cluster to go offline without impacting the availability of the data.

Synchronization of the partition's main index is critical to ensuring that requests can be handled by any server in the cluster. Updates to the main index are replicated to all servers in the cluster. This ensures that each index maintains a reference to all tables currently available in the cluster. The future-dating of updates to the main index across servers allows updates to be propagated to all nodes in the cluster before they take effect. If a main index is not available for synchronizing, the server is taken offline and is not included as part of future requests until it has been restored and re-added to the cluster.

Cataloging of new data thus becomes a simple lookup in the partition's main index on any of the servers in the cluster to determine where the data should be written. The manner in which data is distributed (e.g., the location and number of copies), within the partition is configurable. A setting in the partition specifies the number of servers on which the data is to be catalogued (including allowing for a setting of one which results in a single copy being stored with no redundancy), as well as the number of geographic locations that should be supported.

By providing a capability to redundantly catalog data, an Adaptive Partition allows any of the servers in its cluster to go down without impacting the partition's ability to service data requests. When the cluster detects that a table (or tables), is offline, it flags them in the main index and from that point forward they are not considered valid for cataloging.

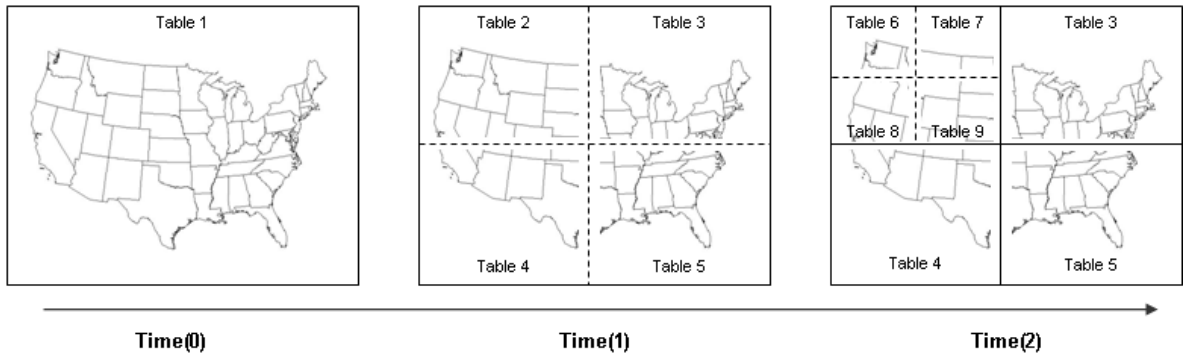

**Figure 2: GTPI and Active Tables in Partition**

Conversely, when a server that has been out of service is restored to the cluster, or when a new server is added, a request may be sent to the Adaptive Partition to force the tables in the partition to age out (i.e., close). As the tables are aged out and new tables are created to take their place, the newly added server is considered as a possible destination when determining how to distribute the new live tables. Closed tables in the cluster may be replicated or copied on other servers, thereby providing a greater degree of availability for the table being copied.

To better distribute load across servers and to avoid hot spots, locations for new tables are determined using a formula that calculates the relative load for each server in the cluster. If an optimum record count size $C_O$ and age $A_O$ are defined for a table in the partition, then the optimum rate at which a table should grow $R_O$ is $C_O / A_O$. Each table has its own growth rate ($R_A = C_A / A_A$). The load factor a particular table (before splitting) is simply $L_T = R_A / R_O$. A server's load in the cluster is calculated by summing load across tables it manages, or $\sum L_T$.

**STOP's First Adaptive Partition**

While STOP's users focus primarily on their devices, and more specifically devices that they are currently tracking, the more interesting and processing intensive queries tend to center on satisfying requests for data that ask what devices were at a particular location for a particular period of time. STOP's first Adaptive Partition addressed this query using a main index that re-distributes data based on time and a device's geographic location but does so irrespective of device (i.e., return all devices that were at a particular location at a point in time).

The main index of this geo-time spatial Adaptive Partition is referred to as the Global Track Point Index (GTPI). Each entry in the GTPI includes a geographic latitude/longitude range (setting up a bounding box geographically) as well as a start and end date for when the data was reported. As data flows into the Adaptive Partition and certain tables reach an optimum record level quicker than expected, they are split and four new tables are created to accommodate a higher transaction throughput rate.

Figure 2 provides a visual representation of how the GTPI and its Adaptive Partition evolve as data is cataloged. Observation shows that over time the partition's table distribution aligns very closely with the characteristics of inbound device data. Geographic locations that have a higher concentration of reporting devices most often have GTPI records that split several times to accommodate the volume of data.

Another important operational benefit gained from Adaptive Partitioning is the ability to more easily archive data as a result of tables referenced by the GTPI being replaced and closed to inbound transactions. Once a data table exceeds the service level agreements for keeping it online, it is archived and dropped from the partition, and the GTPI is updated. At a customer's request, the process is reversed to add data back into the Adaptive Partition.

**Future Work**

There are several opportunities to improve Adaptive Partitioning. First, the solution discussed in this article assumes an even distribution of hardware, network latency (especially when accessing over a WAN), and distribution and data access patterns (i.e., current data is more frequently requested by users).

An issue with the first of these assumptions is the inequalities in hardware and network latency. These inequalities offer an opportunity to focus on refining the clustering solution with respect to Adaptive Partitioning. The inequalities in the performance of nodes within the cluster could be addressed by enhancing the self-evolving nature of the adaptive partition itself. For example, query scrambling inbound requests [AFU 98] and routing them to nodes that can better service them could help eliminate cluster hot spots.

At the same time, STOP is also investigating various cloud offerings that could take advantage of the shared-nothing architecture of Adaptive Partitions and simplify or eliminate certain complexities in the architecture. For example, migrating to Amazon Web Service's Relational Database Service (RDS) would offer the ability to redundantly store data and possibly remove the need for clustering an Adaptive Partition.

**Conclusion**

Adaptive Partitioning resulted from the belief that it could address the transactional latency that was being experienced with STOP's production track point data store. The growing complexity of requests for data coupled with cataloging data in real-time resulted in poor response times and, in some severe cases, a backlog in the cataloging process. Processing hot spots also appeared, especially with more current data, further impacting the data store's performance.

Evaluating the worst offending requests for data led to a discovery that queries focusing on a particular geographic region irrespective of device were the most problematic. STOP's first use of Adaptive Partitioning therefore was the creation of a partition that had both geographic and time-spatial indexing components across all devices. As track points are cataloged, they are distributed based on their geographic location and the date they are reported to the partition. Areas with a higher throughput of data have more tables, while, conversely, other less active locations do not split as often resulting in fewer tables that age out due to inactivity.

Architecturally, the shared-nothing distributed approach, although built on top of an RDBMS, was based on concepts from the NoSQL movement. The more robust feature set of a traditional RDBMS was sacrificed for the simplicity and scalability of being able to distribute data across a series of smaller MySQL databases and tables. Adaptive Partitioning uniquely allows horizontal scaling on locally sufficient "commodity" servers with its ability to evolve as the data set it is servicing changes. As the data characteristics change (e.g., a particular geographic region increases its device saturation), the partition is able to compensate by adjusting its partitioning scheme.